\documentclass[pra,aps,twocolumn,showpacs,10pt]{revtex4-1}
\usepackage{amsmath,amssymb,graphicx,amsthm,dsfont,bm,color,ulem}

\begin{document}

\title{Entangling  non planar  molecules via inversion doublet transition\\ with negligible spontaneous emission}

\author{Isabel Gonzalo}
\affiliation{Departamento de \'{O}ptica, Facultad de Ciencias
F\'{\i}sicas, Universidad Complutense de Madrid, 28040 Madrid, Spain}
\email{igonzalo@fis.ucm.es}

\author{Miguel A. Ant\'{o}n}
\affiliation{Departamento de \'{O}ptica, Facultad de \'{O}ptica y Optometr\'{\i}a, Universidad
Complutense de Madrid,  28037 Madrid, Spain}
\email{antonm@fis.ucm.es}

\begin{abstract}

We analyze theoretically  the entanglement between two non-planar and light identical molecules (e.g., pyramidal as $NH_3$) that present inversion doubling due to the internal spatial inversion of their nuclear conformations by tunneling.
The peculiarity of this system  lies in the simplicity of this type of molecular system in which two near  levels can be connected by allowed electric dipole transition with considerable value of the dipole moment transition and negligible spontaneous emission because the transition is in the microwave or far-infrared range.
These properties  give place to entanglement states oscillating by free evolution with frequency determined by the dipole-dipole interaction  and negligible spontaneous decay, which allows to consider an efficient quantum Zeno effect by frequent measurements of one of the entangled states. If the molecules are initially both in the upper (or lower) eigenstate, the system  evolves under
 an external radiation field, which can induce oscillations of the generated entangled states, with frequency of the order of the Rabi frequency of the field. For a certain detuning,  a symmetric entangled state, eigenstate of the collective system can be populated, and given its negligible spontaneous emission, could be
 maintained for a time only limited by external decoherence processes which could be minimized.
Although the data used are those of the $NH_3$ molecule, other molecules could  present the same advantageous features.

\end{abstract}

\pacs{Entanglement, two-qubits, inversion doubling, non-planar molecules,  dipole-dipole interaction, $NH_3$  }

\maketitle

\section{Introduction}

Quantum entanglement is an essential feature of quantum mechanics and a fundamental resource for quantum computing \cite{Mermin,DiVincenzo,BOOK1,BOOK2,REV1} and quantum information \cite{Bennett2,Gisin,INF1}.
However, entanglement is fragile because of decoherence induced by the environment and it is essential  to maintain quantum entanglement for  long enough time for many applications of interest (e.g., see the work of Horodecki \cite{Horodecki} and references therein).
Since an atom seems good candidate as quantum bit (qubit), several schemes for the generation of entangled states between two atoms have
been proposed, the entanglement being achieved by the coupling of the atoms to a common  reservoir, or dipole-dipole interaction generated in electronic transitions including spontaneous emission, the atoms being eventually  inserted in cavities or/and driven by external radiation fields \cite{Ficek04,Ficek06,Ficek10,Patrick10,Julsgaard12,Bashkirov14,Zhao17} (we quote only some works among many others).

For more than a decade there has been an increasing interest in the possibilities that molecules offer as candidates for quantum bits \cite{Charron07,MOL1,MOL2,MOL3,MOL4}.
For example, polar molecules trapped in an optical lattice can be good candidates for quantum bits.
In this type of lattice, the electric dipole moment of cold or ultracold polar molecules oriented by an external electric field may serve as  quantum bit, and entanglement between qubits is due to the dipole-dipole
interaction  between molecules \cite{Kotochigova06,MOL44,MOL5,MOL6}.
Technical advances make possible to prepare oriented \cite{Grishanin05,Babilotte16} and cold and ultracold molecules \cite{Carr09,Jin12,Park17,COL1} with promising results.
There have been considerable advances in improving coherence times in molecules, obtaining coherence times such as  0.7 ms at 10 K  and 1 $\mu$s at room temperature \cite{Bader14,Ferrando-Soria16}.
Some of the advantages that molecules offer are their internal degrees of freedom and polarities.
The interest of their entanglement not only lies in quantum information and quantum computation but also in processes involving several molecules.

Here we address the entanglement between two identical light non-planar molecules (as $NH_3$, for example) by means of their internal degree of freedom of spatial inversion of their nuclear conformation by tunneling, which corresponds to the well known splitting of their rovibrational levels, named inversion doubling.
This type of  molecules is modeled by a symmetric electronic double well potential in which the molecule oscillates by tunneling through a potential barrier between two degenerate equilibrium conformations, each one being mirror image of the other and corresponding to each minimum of the double well (see Fig. 1(a)).
A well known molecule representative of this behaviour is the pyramidal molecule $NH_3$, the mirror plane being a plane parallel to that determined by the three hydrogens.
This model is also applied to describe the two enantiomers of a chiral molecule, or of a non-planar molecule with axial chirality as $H_2O_2$ where the mirror plane contains the $O-O$ axis.

As is well known, the frequency  of the tunneling oscillation is half the resonant frequency $\omega_0$ between the two levels of the inversion doubling, in the microwave or far infrared range. Due to this low frequency range, the spontaneous emission from the upper level is very small or negligible as we shall see later. The separation between these two levels depends on the height of the barrier and the mass of the molecule and is extremely  small or practically null in heavy molecules in which it is impossible to  observe it because of the extremely long tunneling times involved; decoherence dominates and coherent tunneling is  destroyed o suppressed  by  decoherence induced by the environment, the localized pyramidal or chiral conformations becoming stable states \cite{Zureck,Isabel,Jonalassinio,IsaBar2011,Coles}.
This is the case of $PH_3$, for example, and  chiral molecules, in general.
However, in light molecules whose tunneling times are short enough,  the coherence of  tunneling  can be maintained for some time under appropriate conditions. This is the case of the so much studied  molecule $NH_3$ and also of the above mentioned molecule $H_2O_2$, with $\omega_0 \sim 10^{11}-10^{12}$ rad s$^{-1} $\cite{Quack}, for which tunneling coherence could be preserved for times of the order of $\mu$s at low enough pressure and temperature \cite{Isabel}, or  even for longer time according to technical advances \cite{Bader14,Ferrando-Soria16,Carr09,Jin12,Park17}.

\begin{figure}
  \centering
  \includegraphics[width=8.5cm]{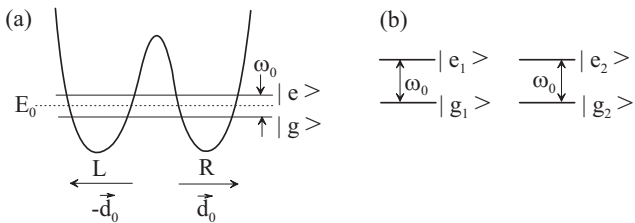}
   \caption{\label{Figure1} (a) Scheme of the inversion doubling for a non planar molecule (see the text). L and R are  the localized conformations, each one mirror image of the other, with the same average energy $E_0$ and, in the case of $NH_3$, with opposite electric dipole moments of absolute value $d_0$.   The eigenstates $|g\rangle$ and $|e\rangle$ of the doublet  are separated in energy $\hbar \omega_0$. (b) Scheme of two identical molecules, each one modeled as in (a) by a two-level system.}
\end{figure}

 Both molecules $NH_3$ and $H_2O_2$ have  similar advantages concerning tunneling times and dipole moments but
 we focus on the  $NH_3$ molecule because it is a symmetric top and this fact seems  advantageous in order to prepare an oriented sample along the axis of the pyramid, maintaining the rotational states around this axis without affecting the specular reflection mentioned. The $H_2O_2$ molecule, instead, is not a symmetric top and has a more complex level structure than $NH_3$.

 The value of the dipole  moment of $NH_3$ is $d_0 = 1.46$ D \cite{Townes},  directed along the height of the pyramid, oscillating by tunneling between the opposite values $\vec{d}_0$ and $ -\vec{d}_0$ (see Fig. 1(a)). This value is high enough to generate, via dipole-dipole interaction, entanglement at intermolecular distances of several nanometers.
 The frequency $\omega_0$ of the inversion doublet increases as the rotational number increases since  the potential barrier results less higher. For example, for $NH_3$, $\omega_0 =1.5\times 10^{11}$ rad s$^{-1}$ in the lowest rotational state that presents inversion doubling, and $\omega_0=6.78 \times 10^{12}$ rad s$^{-1}$ in the next excited rotational state \cite{Townes}. We recall that the rotational energies are larger (infrared) than inversion doubling (microwave or far infrared).

The aim of this work is to study the entanglement generated by dipole-dipole interaction between two identical molecules of this type, each one modeled as a two-level system  with transition frequency $\omega_0$ (see Fig. 1 (a) (b)). The most important advantage and novelty of this system is that the two levels are connected by
allowed electric dipole transition with a frequency $\omega_0$  low enough to consider negligible spontaneous emission whereas the dipole transition moment is high enough, greater than 1 D.

We have found that if the initial state is not eigenstate of the collective system, it evolves freely generating an entanglement oscillating with frequency given by the dipole-dipole interaction and negligible spontaneous emission, allowing an efficient quantum Zeno effect \cite{Zeno,Zeno2,Zeno3} to preserve the entanglement by frequent enough observations.
If the initial state is  eigenstate of the collective system, such as both molecules in the ground or in the excited state, the system only evolves under an external radiation field.
The generated entanglement oscillates then with frequency close to the Rabi frequency.
With proper detuning of the field, dependent on the dipole-dipole interaction, the symmetric entangled eigenstate of the system is populated, and given its negligible spontaneous emission, it could be preserved, its coherence being only limited by environmental decoherence processes which could be minimized according to technical  advances referred to above.

The article is organized as follows. In the next Sec. 2, we present the model which is, in many aspects, analog to that used in works dealing with two atoms, but now adapted to the peculiarity of the  system considered. In Sec. 3, we  analyze  the free temporal evolution of the entanglement between the two molecules for particular initial states. In Sec. 4, the entanglement is analyzed under a coherent external driving radiation field. Sec. 5 contains the conclusions.

\section{The model}

We consider two identical molecules, each one considered as a two-level system corresponding to the inversion doubling in a double well potential, as shown in Fig. 1(a)(b).
We use the notation of  chiral molecules for the two localized conformations of a single molecule, each conformation being mirror image o the other one, lefthanded (L) and righthanded (R), although in the case of $NH_3$, the two mirror images have
the same conformation (regular pyramid) but opposite electric dipole moments.
These states have the same average energy, $E_0$,  they are not eigenstates and oscillate between them by tunneling through the barrier of the double well.
As is known, the lower and upper level, $|g\rangle$ and $|e\rangle$, of the inversion doubling in a molecule of this type can be expressed respectively by,
   \begin{eqnarray}\label{Basic}
   |g\rangle &=& \frac{1}{\sqrt{2}} (|L\rangle +|R\rangle),   \\
   |e\rangle &=& \frac{1}{\sqrt{2}} (|L\rangle -|R\rangle), \label{Basic2}
   \end{eqnarray}
where $|L\rangle$ and $|R\rangle$ are localized states associated to the L and R conformations respectively (pyramidal in the case of $NH_3$ and eventually axially chiral for  molecules as $H_2O_2$). These states, $|g\rangle, |e\rangle$, are delocalized states between the L and R conformations, have null electric dipole moment  and are eigenstates of the free Hamiltonian of the molecule with respective eigenvalues $E_g$ and $E_e$.

The Hamiltonian of two noninteracting identical molecules is given by
$H_0 = \hbar\frac{\omega_0}{2}(\sigma_z^{(1)}+ \sigma_z^{(2)} )$, where $\sigma_z$ is the Pauli matrix $\sigma_z=\sigma_{ee} -\sigma_{gg} = |e\rangle \langle e| - |g\rangle \langle g| $, the superindex  denoting each one of the two molecules, and $\hbar\omega_0 = E_e - E_g$.
It can be easily seen from the above relations (\ref{Basic}) and (\ref{Basic2})  that if the electric dipole moment of the localized conformations is $d_0 =  \langle R |d |R\rangle= -\langle L |d |L\rangle$, the  transition dipole moment is $\mu_{eg}\equiv \langle e |d |g\rangle = - d_0  $.
Then, for a single molecule, the dipole moment operator in the $\{|e\rangle,|g\rangle\}$  basis reads $d = -d_0 (\sigma_- + \sigma_+)$, where $\sigma_{\pm}$ are the raising to the excited state and lowering to the ground state matrix Pauli operators.
The Hamiltonian of the dipole-dipole interaction between the two molecules, $H_{12}$,  is then proportional to
$d^{(1)} d^{(2)} = d_0^2 (\sigma_+^{(1)} + \sigma_-^{(1)}) ( \sigma_+^{(2)} + \sigma_-^{(2)}  )$. Thus we consider $H_{12}= \hbar J (\sigma_+^{(1)}\sigma_-^{(2)} + \sigma_+^{(2)}\sigma_-^{(1)}  )  $, where
$ J \simeq V/\hbar$, with
$V =2 d_0^2 / (4 \pi \epsilon_0 r^3)$ being the dipole-dipole interaction between the two molecules separated by a distance $r$ and with  the dipole moments parallel between them.

We shall consider eventually an external coherent radiation field of frequency $\omega_l$ resonant o quasi-resonant with the two levels $|e\rangle,|g\rangle$,  the interaction Hamiltonian  field - molecules  being, in the rotating wave approximation,  $H_f = \hbar \Omega[(\sigma_+^{(1)}+ \sigma_+^{(2)})e^{-i\omega_l t}+(\sigma_-^{(1)}+ \sigma_-^{(2)})e^{i\omega_l t}] $, where $\Omega = \frac{\vec \mu_{eg} \cdot \vec E_l}{2\hbar}$, with $E_l$ the electric field amplitude.
If we apply the unitary transformation $U = e^{-i \frac{\omega_l}{2}( \sigma_z^{(1)}+ \sigma_z^{(2)} ) }$ to the total Hamiltonian $H = H_0 +H_{12} + H_f$, it becomes (named again $H$):
       \begin{eqnarray}\label{Hamiltonian}
       H&=& \hbar \frac{\Delta_l}{2} ( \sigma_z^{(1)} + \sigma_z^{(2)} )+ \hbar J (\sigma_+^{(1)}\sigma_-^{(2)} + \sigma_+^{(2)}\sigma_-^{(1)})\nonumber \\
       &+& \hbar \Omega ( \sigma_+^{(1)}+ \sigma_+^{(2)} + \sigma_-^{(1)}+ \sigma_-^{(2)}   ),
        \end{eqnarray}
where $\Delta_l =\omega_0 - \omega_l $.
The spontaneous emission from level $|e\rangle$ to level $|g\rangle$ of the same molecule or of the other one is neglected because of the so small value of $\omega_0$ that gives a value for the Einstein coefficient $A = |\mu_{eg}|^2 \omega_0^3 / (3 \pi \hbar \epsilon_0 c^3)$ extremely small, of the order of $10^{-7}$ s$^{-1}$ or $10^{-2}$ s$^{-1}$ for $\omega_0= 1.5 \times 10^{11}$ or $\omega_0= 6.78 \times 10^{12}$ respectively, thus, the spontaneous decay, contributing in part to the decay of entanglement, can be neglected here.

The temporal evolution of the system is governed by the master equation,
     \begin{equation} \label{master}
      \frac{\partial \rho}{\partial t}   = -\frac{i}{\hbar}[H, \rho]+  {\cal L}_d (\rho),
     \end{equation}
where $\rho$ is the density matrix of the state and ${\cal L}_d (\rho)$ is a  pure dephasing Lindblad term that accounts for the loss of coherence of the superposition states (\ref{Basic}) and (\ref{Basic2}). This loss of coherence is equivalent to the loss of coherence of tunneling between the $L$ and $R$ conformations of the molecule, and can be  originated by elastic collisions from the environment. The dephasing term is then given by,
     \begin{eqnarray}
   {\cal L}_d (\rho)&=& -\frac{\gamma}{4}[(\sigma_z^{(1)}\sigma_z^{(1)} \rho + \rho \sigma_z^{(1)}\sigma_z^{(1)} - 2\sigma_z^{(1)} \rho \sigma_z^{(1)} ) \nonumber\\
  &+&
  (\sigma_z^{(2)}\sigma_z^{(2)} \rho + \rho \sigma_z^{(2)}\sigma_z^{(2)} - 2\sigma_z^{(2)} \rho \sigma_z^{(2)} )],
   \end{eqnarray}
where the constant $\gamma$ would be of the order of the inverse of the average time between elastic collisions and can be of the order of $\mu$s or even lower for low enough temperatures and densities. At temperatures of the order or slightly below 1 K, the $kT$ energy (k being the Boltzmann constant) does not reach the separation energy $\hbar \omega_0$ of the inversion doubling here considered and an exchange of energy (inelastic collision) is less probable than a mere change in the phase of the states (elastic collision). Here we assume  cold ($ 10^{-3}< T < 1$ K) molecules, so that $kT$ could be lower than the separation between the levels considered.
 If $kT$ is greater than the energy separation of the levels, it is still possible for low enough pressure that the average time between collisions be much longer than tunneling time \cite{Isabel} (and even than the interaction time $J^{-1}$ here considered between molecules). In this case, coherent tunneling can be maintained in some extent, as occurs in a gas of $NH_3$ under normal conditions of pressure and temperature, in which case the inversion doubling transition can still be observed \cite{Jonalassinio}; however the system would be properly described in this case including a dissipative environment. In our case, we restrict ourselves to a pure dephasing contribution due to elastic collisions since inelastic ones are considered much less probable for cold molecules, as said above.

The density matrix elements are calculated in the collective bare basis using the following notation,
     \begin{eqnarray}\label{Basis}
    |1\rangle &\equiv& |g_1, g_2\rangle, \nonumber\\
    |2\rangle &\equiv& |g_1, e_2\rangle, \nonumber\\
     |3\rangle &\equiv& |e_1, g_2\rangle, \nonumber\\
      |4\rangle &\equiv& |e_1, e_2\rangle,
    \end{eqnarray}
where the subindex denotes each one of the two molecules.
The differential  equations for the density matrix elements are then given by:
     \begin{eqnarray} \label{equations}
 \dot{\rho}_{11} &=& -i \Omega(\rho_{31}-\rho_{13}+\rho_{21}-\rho_{12} ), \nonumber\\
 \dot{\rho}_{22} &=&-i\Omega(\rho_{12}-\rho_{21}+\rho_{42}-\rho_{24})-iJ(\rho_{32}-\rho_{23}), \nonumber\\
\dot{\rho}_{33} &=& -i \Omega(\rho_{31}-\rho_{13}+\rho_{34}-\rho_{43})-iJ(\rho_{32}-\rho_{23})   ,  \nonumber \\
\dot{\rho}_{44} &=& -\dot\rho_{11}-\dot\rho_{22}-\dot\rho_{33}, \nonumber \\
\dot{\rho}_{12} &=&  i\Delta_l \rho_{12} - i\Omega (\rho_{22}-\rho_{11} +\rho_{32} -\rho_{14})  +iJ\rho_{13} - \gamma \rho_{12}, \nonumber\\
\dot{\rho}_{13} &=& i \Delta_l \rho_{13} - i\Omega (\rho_{33}-\rho_{11} +\rho_{23} -\rho_{14}) + iJ \rho_{12}- \gamma \rho_{13}, \nonumber\\
\dot{\rho}_{14} &=& i2 \Delta_l \rho_{14}  - i\Omega(\rho_{34}+\rho_{24}-\rho_{12}-\rho_{13})- 2 \gamma \rho_{14}, \nonumber\\
 \dot{\rho}_{23}&=& -i \Omega(\rho_{13}+\rho_{43}-\rho_{24}-\rho_{21}) -i J(\rho_{33}-\rho_{22})-2 \gamma \rho_{23},\nonumber\\
 \dot{\rho}_{24} &=& i \Delta_l \rho_{24} + i\Omega(\rho_{22}+ \rho_{23}-\rho_{44}-\rho_{14}) - iJ \rho_{34}-\gamma \rho_{24}, \nonumber\\
\dot{\rho}_{34} &=& i \Delta_l \rho_{34} + i\Omega(\rho_{33}+\rho_{32}-\rho_{44}-\rho_{14}) - iJ\rho_{24}- \gamma \rho_{34}.
     \end{eqnarray}

In the case that there is not  applied radiation field, i.e., $\Omega=0$, the detuning $\Delta_l$ must be replaced with $\omega_0$.
In this case, the eigenstates of the collective system are well known \cite{Ficek04,Liao10} and are shown in Fig. 2.
\begin{figure}
   \centering
   \includegraphics[width=4.5cm]{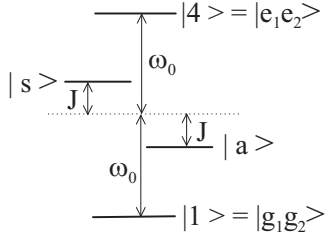}
   \caption{\label{Figure2} Eigenstates of the collective system. The symmetric entangled state $|s\rangle \equiv\frac{1}{\sqrt{2}}(|2\rangle+|3\rangle)= \frac{1}{\sqrt{2}}( |g_1e_2\rangle + |e_1g_2\rangle )   $  and the antisymmetric entangled state $|a\rangle \equiv \frac{1}{\sqrt{2}}(|2\rangle-|3\rangle)= \frac{1}{\sqrt{2}}( |g_1e_2\rangle - |e_1g_2\rangle )  $ are separated by the energy $2\hbar J$.}
 \end{figure}

The antisymmetric entangled eigenstate $|a\rangle \equiv \frac{1}{\sqrt{2}}(|2\rangle-|3\rangle) $ is not connected via electrical dipole transition with the other levels, while the symmetric state $|s\rangle \equiv \frac{1}{\sqrt{2}}(|2\rangle+|3\rangle) $ can be connected by this via with the highest $|4\rangle$ and lowest $|1\rangle$ collective levels.
The energy separation between the entangled states is $2\hbar J$. We shall consider the  $J$  value lower enough than $\omega_0$ so that tunneling does not become altered essentially. The value of J depends on the distance between the two molecules and their orientation. We shall then consider distances between the molecules of the order of $r \sim 10$ nm which gives $ J =V/\hbar \sim 4\times 10^{9}$ s$^{-1}$. The distance considered is much lower than the wavelength associated to $\omega_0$, about 1 cm if $\omega_0= 1.5 \times 10^{11}$ s$^{-1}$.

Since the initial conditions of the molecules could be the localized conformations $L$ or $R$, it can be useful to express the following entangled states not only as superpositions of the separable states of the bare basis (Eqs. (\ref{Basis})) but also as superpositions of the separable states in terms of the localized  $|L\rangle$ and $|R\rangle$ states, by using Eqs. (\ref{Basic}) and (\ref{Basic2}), as follows:
      \begin{eqnarray}\label{Entangled}
     |a\rangle&\equiv& \frac{1}{\sqrt{2}}(|2\rangle - |3\rangle) = \frac{1}{\sqrt{2}} (|L_1R_2\rangle -|R_1L_2\rangle),\nonumber \\
     |s\rangle&\equiv& \frac{1}{\sqrt{2}}(|2\rangle + |3\rangle) = \frac{1}{\sqrt{2}} (|L_1L_2\rangle -|R_1R_2\rangle),\nonumber \\
     |p\rangle&\equiv& \frac{1}{\sqrt{2}}(|1\rangle + |4\rangle) = \frac{1}{\sqrt{2}} (|L_1L_2\rangle +|R_1R_2\rangle),\nonumber \\
     |q\rangle&\equiv& \frac{1}{\sqrt{2}}(|1\rangle - |4\rangle) = \frac{1}{\sqrt{2}} (|L_1R_2\rangle +|R_1L_2\rangle).
    \end{eqnarray}
If we call $\vec v_E$  the vector formed by the entangled states, and $ \vec v$ the vector formed by the separable states of Eqs. (\ref{Basis}),
we see they are related between them as $\vec v_E = M \vec v $, with $M$ the corresponding transformation matrix, and in consequence, the density matrix in the entangled states basis, $\rho_E$, is given by $\rho_E = M \rho M^{-1}$, where $\rho$ is the density matrix expressed in Eqs. (\ref{equations}). This relation between the density matrices is useful to analyze the time evolution of  the populations and coherences of the entangled states.

The entanglement between the two molecules will be analyzed by means of the Wootters entanglement measure \cite{Wootters}, the concurrence C, defined as
$C= max (0, \sqrt{\lambda_1}-\sqrt{\lambda_2}-\sqrt{\lambda_3}-\sqrt{\lambda_4} )$, where $\lambda_1, \lambda_2, \lambda_3, \lambda_4$ are the eigenvalues, in decreasing ordered, of the matrix $\rho \tilde{\rho}$, where $\tilde{\rho} = (\sigma_y^{(1)} \otimes \sigma_y^{(2)}) \rho^* (\sigma_y^{(1)} \otimes \sigma_y^{(2)}) $ with $\sigma_y$  the $y$-Pauli matrix. The value of $C$ is between 0 and 1, the unity corresponding to maximum entanglement. For the chosen basis of separable states (\ref{Basis}) we have,
  \begin{equation}\label{sigma}
  \sigma_y^{(1)} \otimes \sigma_y^{(2)} = \left( \begin{array}{cccc}
                                              0 & 0 & 0 & -1 \\
                                              0 & 0 & 1 & 0 \\
                                              0 & 1 & 0 & 0 \\
                                              -1 & 0 & 0 & 0
                                            \end{array}\right).
  \end{equation}

\section{Entanglement without driving radiation field}

First we consider that there is not  applied radiation field, i.e., $\Omega=0$, hence,  $\Delta_l$ is replaced with $\omega_0$ in the above equations (\ref{equations}). We use the following data of  the  $NH_3$ molecule:  frequency of the inversion doubling $\omega_0 = 1.5\times 10^{11}$ rad s$^{-1}$,  electric dipole moment of the pyramidal conformation $d_0=1.46$ D  (the same as the transition dipole moment, $|\mu_{eg}|$). We consider $J= 4\times 10^{9}$ s$^{-1}$ and a coherence decay by dephasing given by  $\gamma =10^6$ s$^{-1}$,   although smaller values (longer molecular time coherence) seem to be possible according to promising results referred in the Introduction.
Depending on the initial conditions we have different cases:

A) In this case we consider that the initial state is the separable state in which one of the molecules is in the excited state $|e\rangle$ and the other one in the ground state $|g\rangle$, such as the state $|3\rangle \equiv |e_1 g_2\rangle$,  of interest in processes of excitation transfer.
This case differs from the excitation transfer process with electronic levels and then appreciable spontaneous emission, studied with detail by other authors (e.g., Ficek \cite{Ficek04} and Liao \cite{Liao10}).

As can be easily  deduced from Eqs. (\ref{equations}) in our simple case, the state $|3\rangle$ evolves in time giving place to oscillations of the populations $\rho_{22}$ and $\rho_{33}$ (the only ones involved) as $\sin^2 (Jt)$ and $\cos^2 (Jt)$ respectively,  while the concurrence oscillates as $|\cos (2Jt)|$.

 By analyzing the density matrix elements at times where the concurrence has maximum values, it can be seen that the maxima of the concurrence $C$ correspond alternatively to each one of the entangled states $|f\rangle \equiv\frac{1}{\sqrt{2}}(|2\rangle + i |3\rangle)$ and $|k\rangle \equiv\frac{1}{\sqrt{2}}(|2\rangle - i |3\rangle)$ whose populations oscillate  with frequency $J$.  The population of the entangled state $|f\rangle$, $\rho_{ff}$, is represented together with the concurrence in Fig. 3 (a).
The decay due to  the $\gamma$ value considered results obviously imperceptible in the range of nanoseconds determined by the $J$ value. Even in the case of considering a larger value of $\gamma$ such as $\gamma=10^{7}$ s$^{-1}$, the maximum of concurrence would descend from $C=1$ to $C= 0.9$ after 20 ns from the initial time.
The population of the other entangled state $|k\rangle$, not represented in the figure,  reaches maximum population when $\rho_{ff}=0$. If $J$ is lower than the value considered (more separation between the molecules), the oscillations become slower.
 \begin{figure}
  \centering
  \includegraphics[width=8cm]{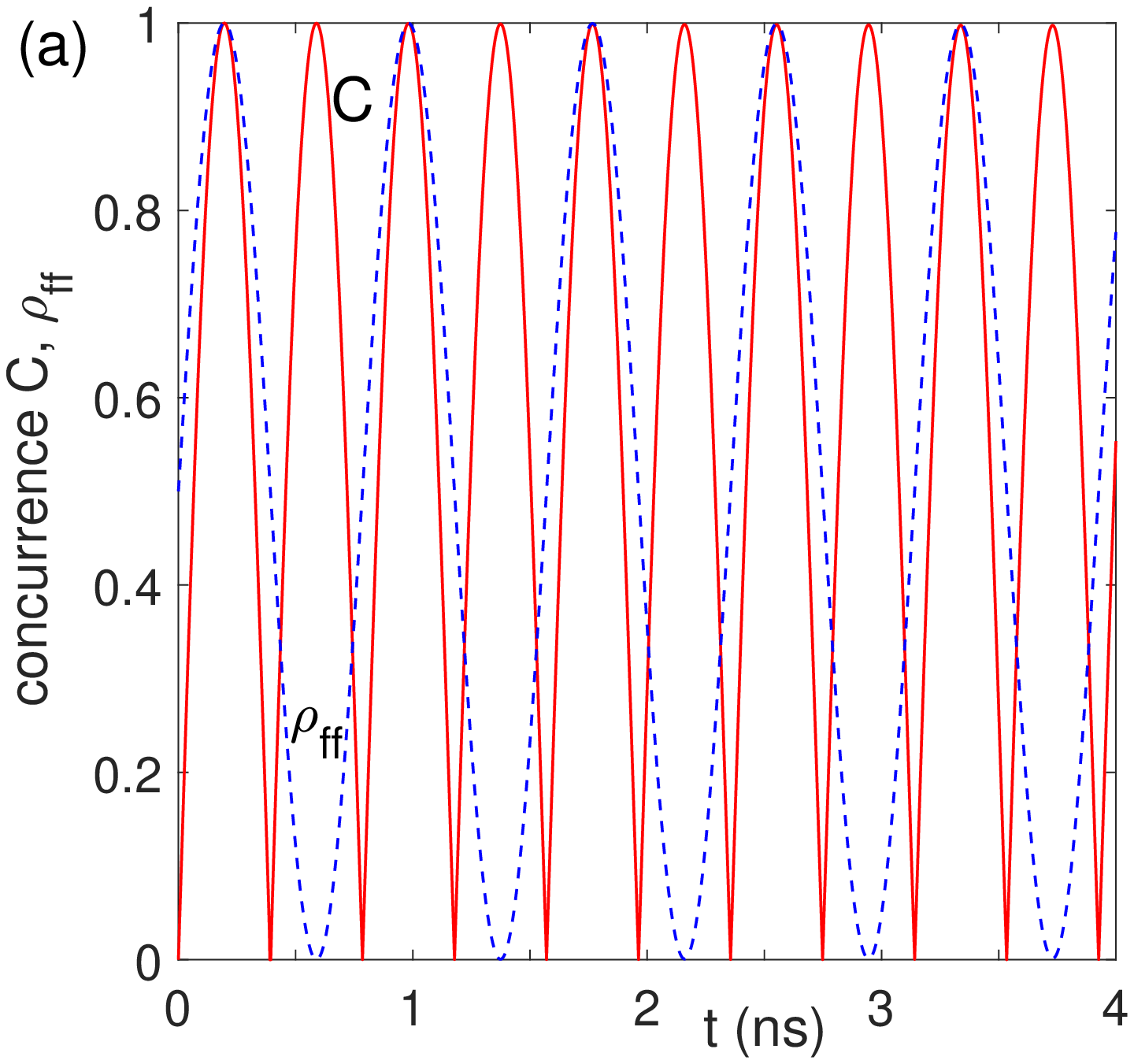}
  \includegraphics[width=8cm] {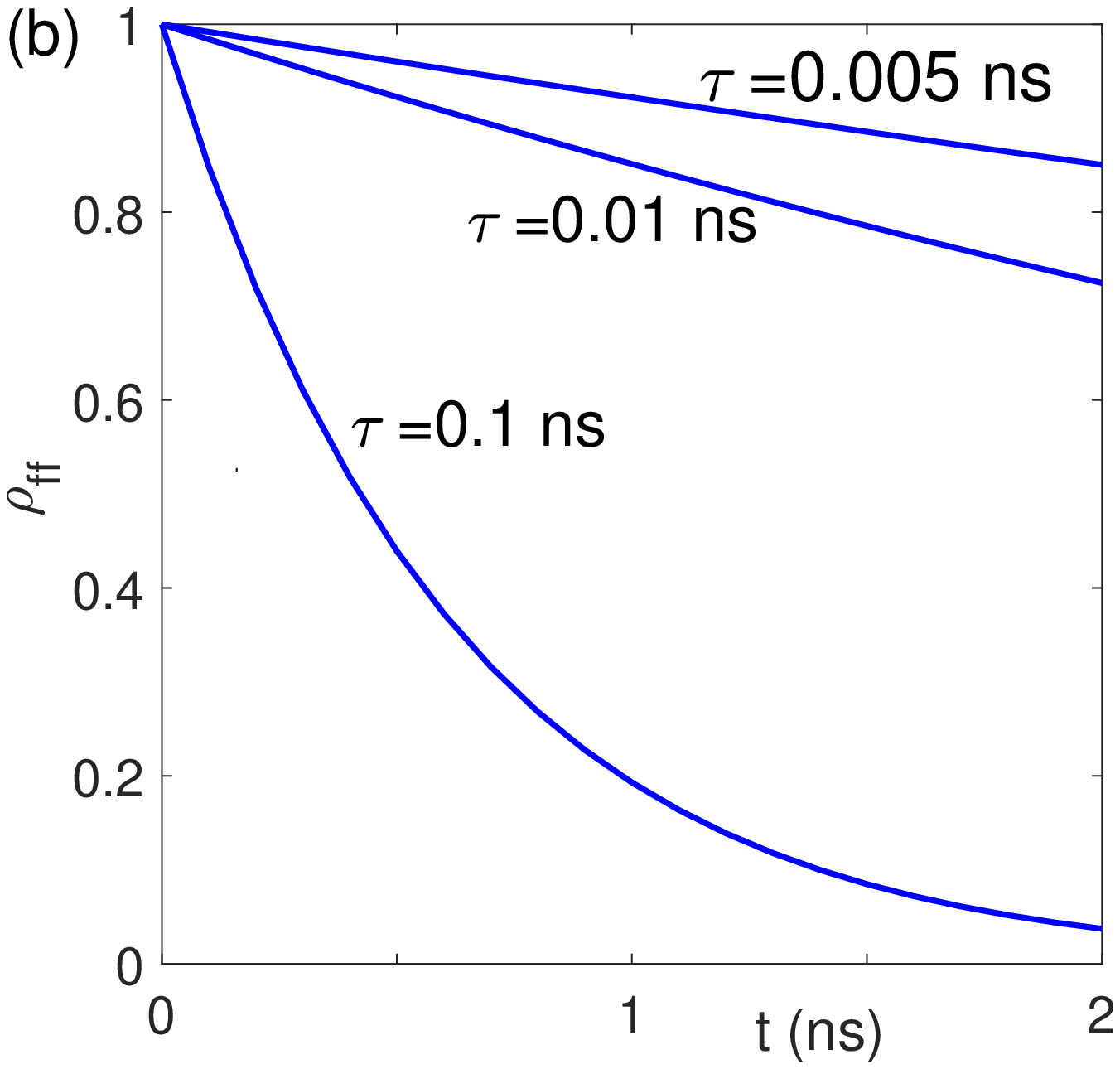}
  \caption{\label{Figure3} (a) Temporal evolution of  concurrence C (solid line) and population $\rho_{ff}$ of the entangled state $|f\rangle \equiv\frac{1}{\sqrt{2}}(|2\rangle + i |3\rangle)$ (dashed line), for the initial state $|3\rangle \equiv |e_1g_2\rangle$, $\omega_0=1.5\times 10^{11}$ rad s$^{-1}$,  $J=4\times 10^{9}$ s$^{-1}$ and $\gamma=10^{6}$ s$^{-1}$.
  (b) Zeno effect for the probability to find the entangled state $|f\rangle$ by projective measurements of the state at time intervals $\tau = 0.1, \, 0.01$ and $0.005$ ns, for the same data as in (a).}
\end{figure}

Given the practically harmonic temporal dependence of the populations, this case offers an easy possibility to quantum Zeno effect, thus  avoiding or retarding  the temporal evolution of one of the entangled states, $|f\rangle$ or $|k\rangle$, then keeping a high degree of entanglement for a time near the $\mu$s range where the decoherence manifests due to the value of $\gamma$ considered. Quantum Zeno could then be applied at the time when $\rho_{ff}$ or $\rho_{kk}$ are the unity. We choose $\rho_{ff}=1$. From this value, this population evolves practically as $\rho_{ff}=cos^2 (Jt)$ and, as it is well known, it allows quantum Zeno effect \cite{Zeno2} by frequent projective measurements on the state $|f\rangle$. If the measurements are made at short enough time intervals, lower than the so called Zeno time, in our case  $\tau < J^{-1}$, the probability to find the state $|f\rangle$ after $N$ measurements, at the time $T=N\tau \ll 1/\gamma$, is
      \begin{equation}
     (\rho_{ff} (\tau))^N = [cos^2 (J\tau)]^N \simeq (1-J^2\tau^2)^N \simeq e^{-J^2 T^2/ N},
     \end{equation}
which tends to the unity as $N \rightarrow \infty$. An example of Zeno effect in this case is shown in Fig. 3 (b). We can appreciate in this figure that the population of the entangled state $|f\rangle$ does not oscillate but decreases more slowly the more frequent the measurements are.

B) Now we consider the case in which the two molecules are in localized states, $|L\rangle$ or $|R\rangle$, with their dipole moments parallel between them. We recall that for the $NH_3$ molecule there is no physical distinction between L and R conformations.  Since the dipole moment can be oriented in space independent of the chirality conformation L or R in the case of chiral molecules, the initial state  can be $|L_1L_2\rangle$,  $|R_1R_2\rangle$ or $|L_1R_2\rangle$, the results obtained being similar for any of the three  initial states.

We choose, for example, the separable initial state $|L_1L_2\rangle$.
Using the above equations (\ref{Entangled}), this  state can be easily expressed in the  basis (\ref{Basis}) and also as superposition of entangled states, as follows:
      \begin{equation}
     |L_1L_2\rangle= \frac{1}{2}(|1\rangle +|2\rangle +|3\rangle +|4\rangle)= \frac{1}{\sqrt{2}} (|s\rangle + |p\rangle)
     \end{equation}
The temporal evolution of concurrence is shown in Fig. 4 (a) and a detail of the temporal evolution of the populations of the entangled states, $\rho_{aa},\rho_{ss}, \rho_{pp}, \rho_{qq}$, is shown in Fig. 4 (b).

As in the preceding case, the decoherence decay has not visible effect at short times as is this case.
The states $|p \rangle$  and $|q \rangle $ interchange their populations oscillating as $\sin^2 (\omega_0 t)$ and  $\cos^2 (\omega_0 t)$  while the concurrence oscillates as $|\sin(Jt)|$, i.e., with half of the frequency than in the preceding example. For lower values of $J$, a much more slowly variation of the concurrence is obtained, which could be desirable. As expected from the equations (\ref{equations}), the states $|a\rangle$ and $|s\rangle$ keep their  initial populations, zero and $1/2$ respectively.

From the analysis of the density matrix of the evolved state at times in which concurrence takes maximum and minimum values, we see that the  maximum values of the concurrence correspond to an entangled state of the type
$|\Psi\rangle =(1/2) |s\rangle + c_p |p\rangle + c_q |q\rangle$ where $c_p, c_q$ depend on the state of the oscillation.
The zero values of the concurrence correspond  to the separable states $|R_1 R_2\rangle$ and $|L_1L_2\rangle$  in an alternated  way, as expected by symmetry.
\begin{figure}
  \centering
  \includegraphics[width=8cm]{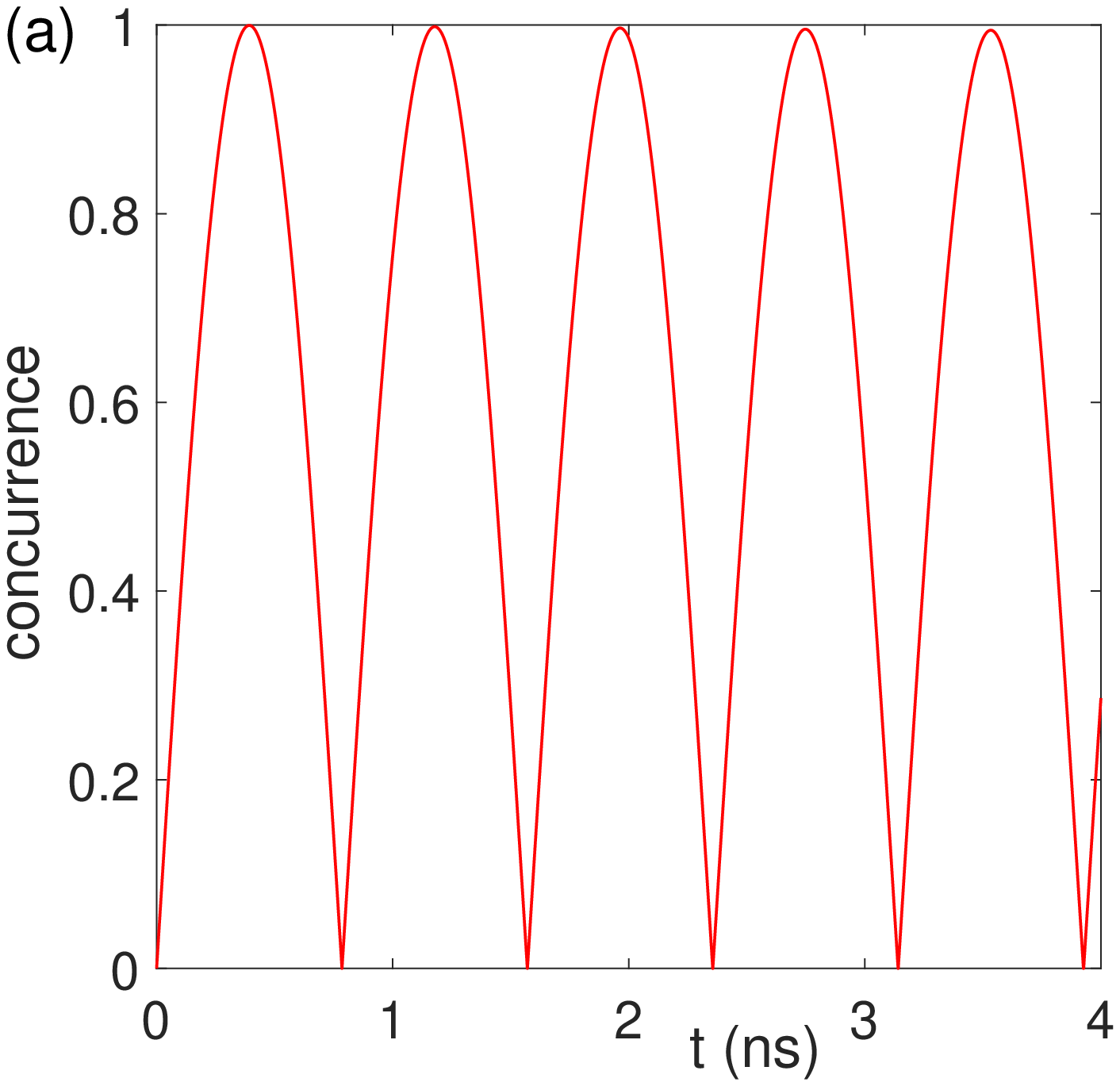}
  \includegraphics[width=8cm]{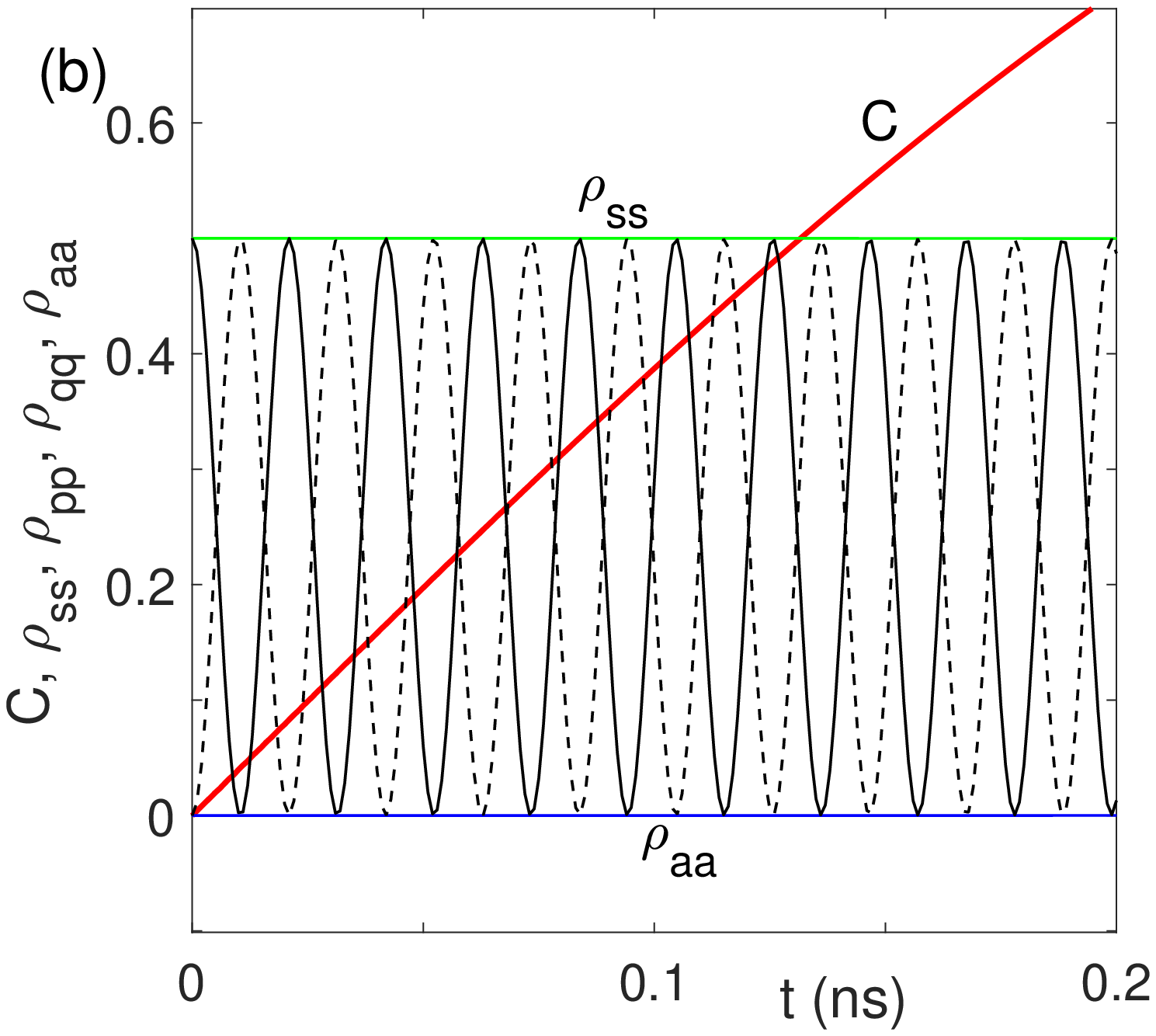}
  \caption{\label{Figure4} Temporal evolution of (a) concurrence  and (b) detail of the populations $\rho_{pp}$ (oscillating solid line), $\rho_{qq}$ (dashed line), $\rho_{ss}$ (horizontal green line), $\rho_{aa}$ (horizontal blue line) and concurrence C (solid red);  for the initial state $|L_1 L_2\rangle$, $\omega_0=1.5\times 10^{11}$ rad s$^{-1}$, $J=4\times 10^{9}$ s$^{-1}$ and $\gamma=10^{6}$ s$^{-1}$.}
\end{figure}

C) Finally, if the initial condition is
     \begin{equation}
      |L_1R_2\rangle= \frac{1}{2}(|1\rangle +|2\rangle -|3\rangle -|4\rangle)= \frac{1}{\sqrt{2}} (|a\rangle + |q\rangle),
     \end{equation}
identical temporal dependence of concurrence as in the preceding case  is found. The temporal dependence of the populations of the entangled states is analogous to the precedent case: $\rho_{pp}, \rho_{qq}$  oscillate in alternated way  and  now $\rho_{ss}=0$, $\rho_{aa}=1/2$. In the entangled states with maximum value of concurrence the contributing states are now $|a\rangle$, $|p\rangle$ and $|q\rangle$.

\section{Entanglement in presence of a driving coherent radiation field}

We consider now a coherent radiation field driving our system initially in the upper, $ |e_1e_2\rangle$, or lower, $ |g_1g_2\rangle$, eigenstate  of the collective system (see Fig. 1(b)). Just this type of state was prepared in the past in the $NH_3$ molecule to obtain the first maser (e.g., \cite{Feynmann}).
Given the negligible spontaneous decay, this type of state can only evolves under the stimulation of a radiation field. We use the same data as in the preceding section and a value for the Rabi frequency of the field lower than the dipole-dipole interaction.
Depending on the detuning with respect to the inversion doubling frequency $\omega_0$, we have the following cases.

\begin{figure*}
\includegraphics*[height=6cm]{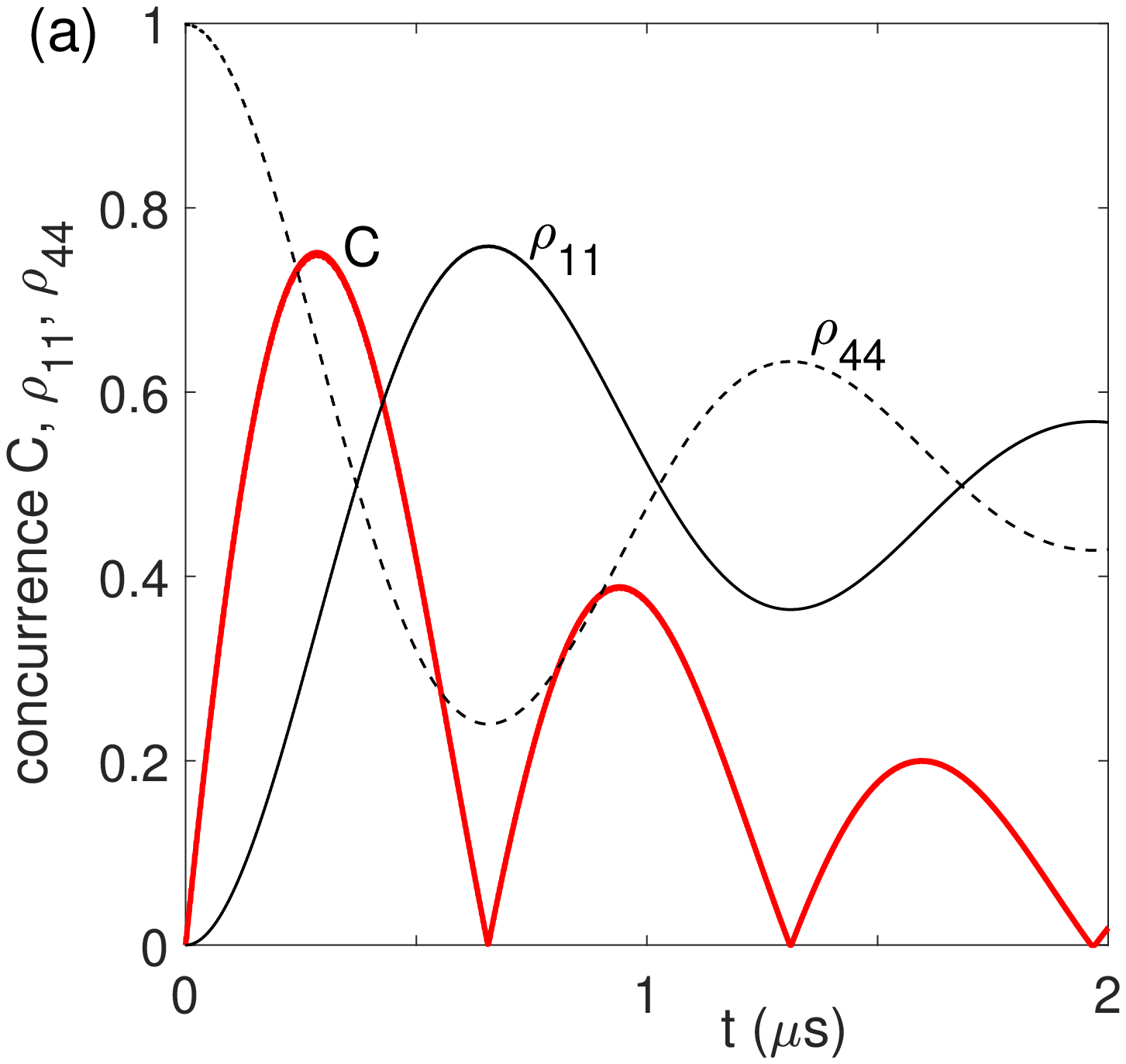}
\includegraphics*[height=6cm]{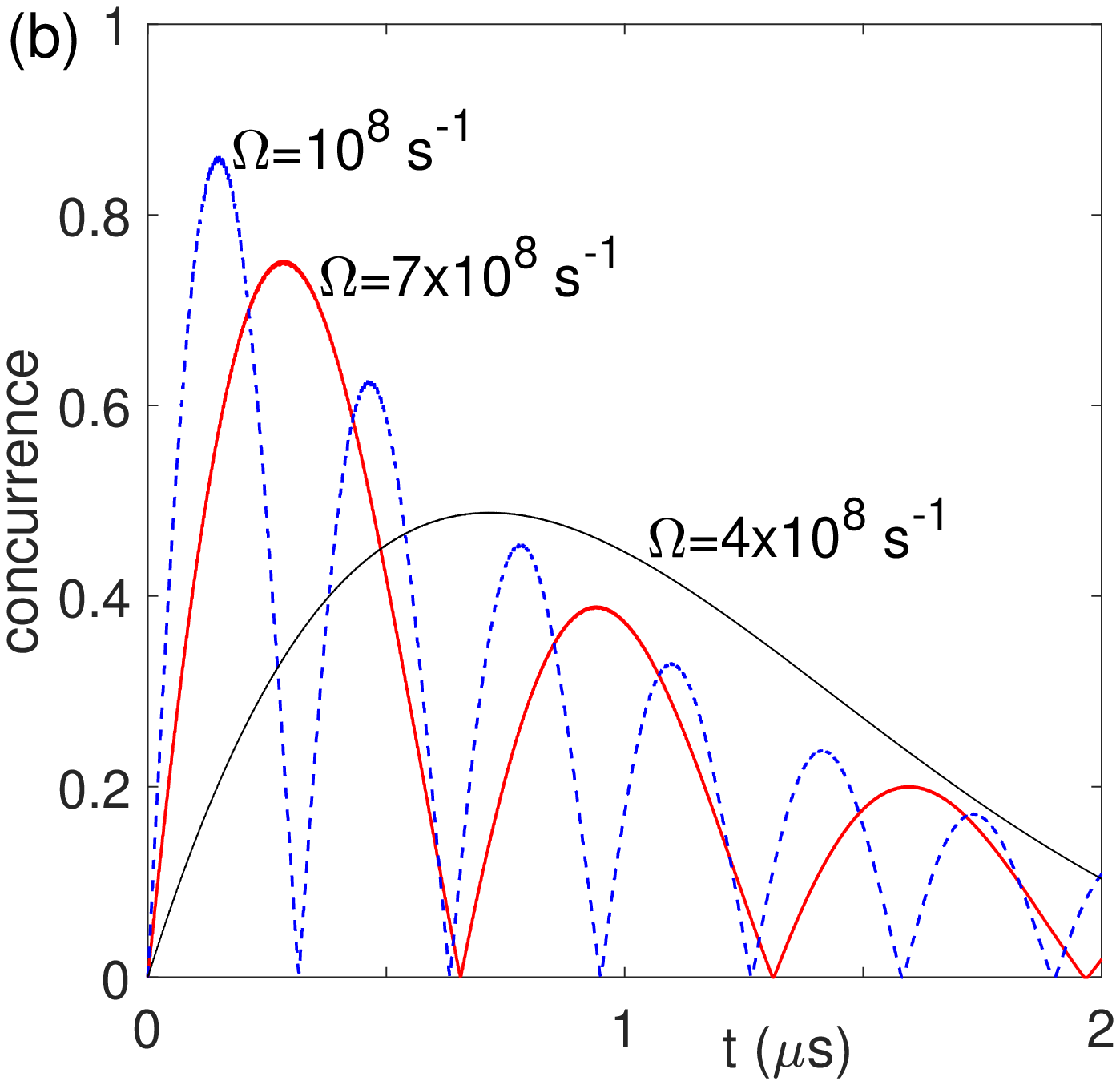}\\
\includegraphics*[height=6cm]{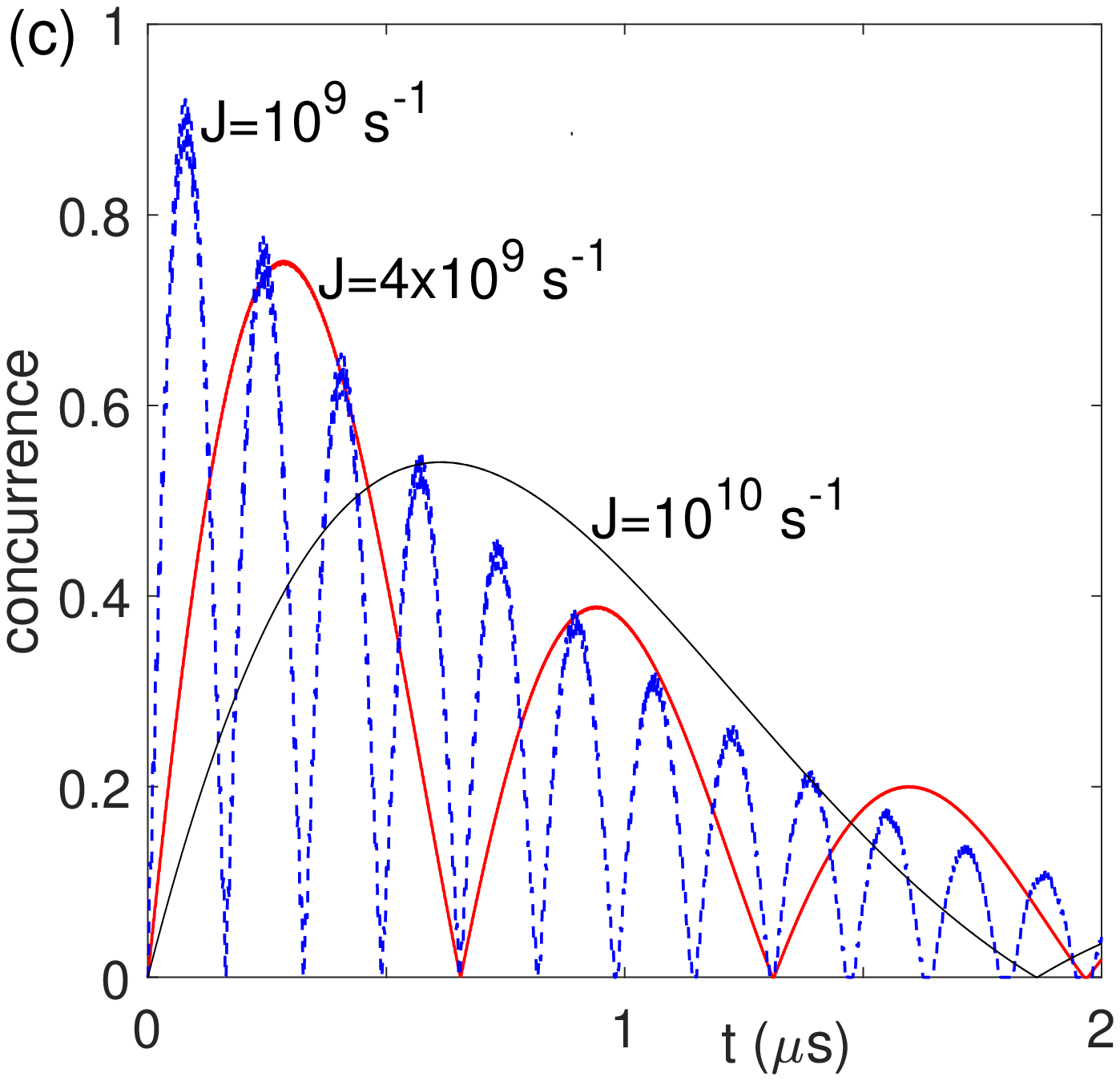}
\includegraphics*[height=6cm]{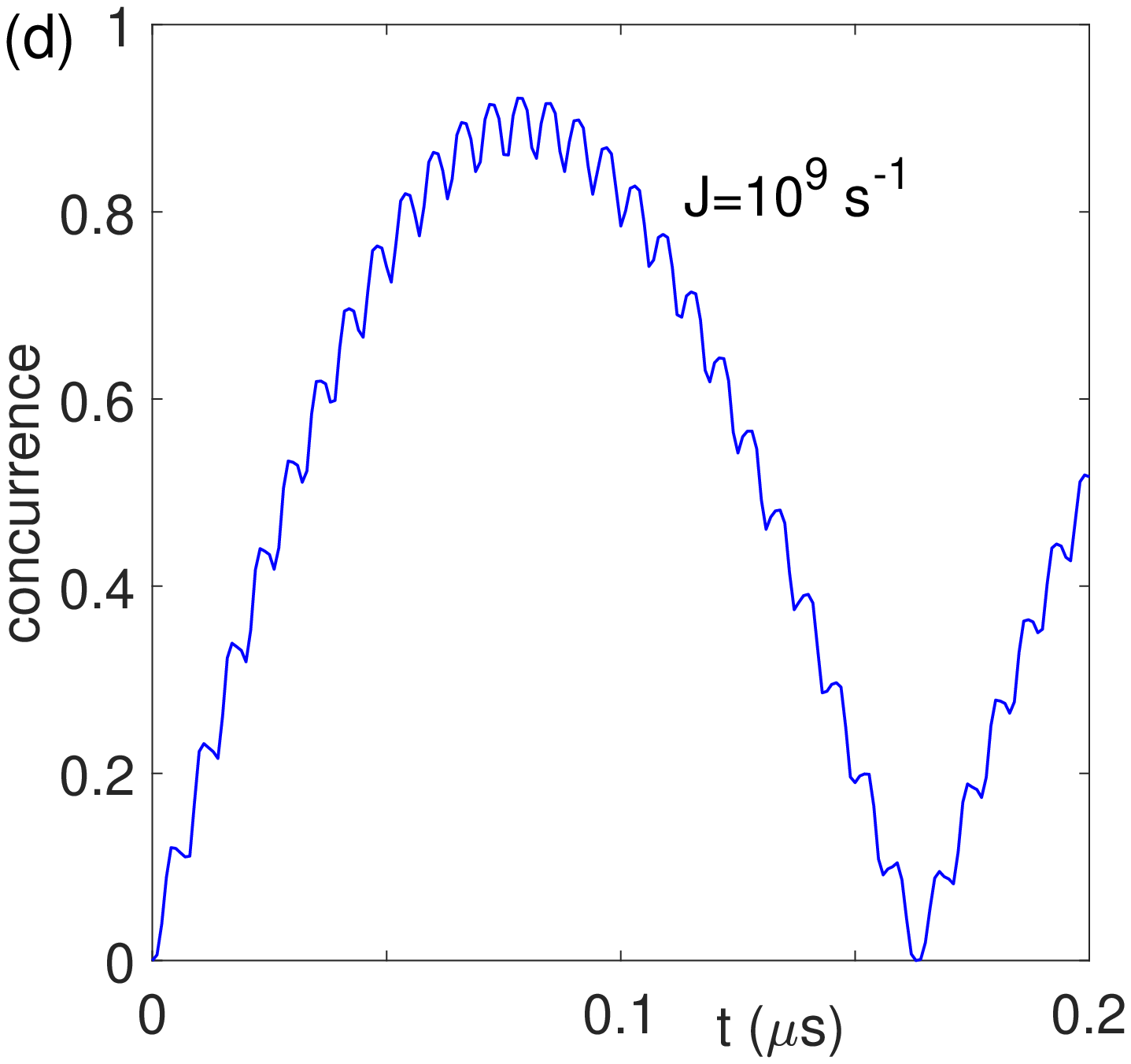}
  \caption{\label{Figure5}For the initial state $|e_1e_2\rangle$, $\Delta_l = 0$, $\omega_0=1.5\times 10^{11}$ rad s$^{-1}$ and $\gamma=10^{6}$ s$^{-1}$, temporal evolution of: (a) concurrence C and populations $\rho_{11}$ and $\rho_{44}$, for $J=4\times 10^{9}$ s$^{-1}$ and $\Omega = 7\times 10^7$ s$^{-1}$; (b) concurrence  for $J=4\times 10^{9}$ s$^{-1}$ and several values of $\Omega$; (c) concurrence for $\Omega=7\times 10^7$ s$^{-1}$   and several values of J; (d) detail of the curve for $\Omega=7\times 10^7$ s$^{-1}$ and $J=10^9$ s$^{-1}$. }
\end{figure*}

A) First we consider resonance condition, $\Delta_l =0$, the initial state being $|4\rangle \equiv |e_1e_2\rangle$, i.e., both molecules in the upper state.
By solving the Eqs. (\ref{equations}) by numerical method we obtain, for the case in which the Rabi frequency is $\Omega = 7\times 10^7$ s$^{-1}$, the temporal evolution of concurrence and populations shown in Fig. 5 (a). It can be appreciated  that the populations $\rho_{11}$ and $\rho_{44}$ oscillate with a period  longer than the Rabi period $2 \pi/\Omega \simeq 10^{-7}$ s.
This effect is due to the dipole-dipole interaction $J$ that occurs during the transition $|4\rangle \leftrightarrow |1\rangle $ and tends to slow down the oscillation, giving place to a  lower effective Rabi frequency.

The effect of different values of $\Omega$ on the concurrence is shown in Fig. 5 (b) where it can be seen how the oscillation of the concurrence  could be controlled by $\Omega$.
Now the decay of the concurrence by decoherence is appreciable since the temporal evolution takes place in a range close to the molecular memory time considered, $\gamma^{-1}=1 \,\mu$s. In the ideal case that no coherence decay occur,   an inspection of the populations at the maxima of the oscillating concurrence allows us to assert that the entangled states corresponding to these maxima are $|\psi\rangle = \frac{1}{\sqrt{2}} (|1\rangle \pm i |4\rangle ) $ in alternated way.
However, the decay due to $\gamma$ makes that the entangled states at the maxima of the concurrence involve the participation of the $|s\rangle$ and $|a\rangle$ states.

In Fig. 5 (c), several cases with different $J$ values, for a given value of $\Omega$ and the same values for the other parameters as in  Fig. 5 (a),  are shown. The greater the value of $J$ the slower the oscillation of the concurrence, as we have said. This behaviour is contrary to what happens when there is no driving field, as  explained in the preceding section. The detail of Fig. 5 (d) shows small oscillations with frequency of the order of J, which become imperceptible for higher J as occurs in the curves of Fig. 5 (c).

The same results as shown in Fig. 5  are found if the initial state is in the ground state $|1\rangle=|g_1g_2\rangle$, as was expected since there is not spontaneous emission in the state $|e_1e_2\rangle$.

B) We analyze now the effect of the detuning, $\Delta_l = \omega_0 -\omega_l \neq 0$. If the initial state is the upper state $|e_1e_2\rangle$, and the detuning is $\Delta_l=J$, we see in Fig. 2 that a resonance is established between the upper level and the symmetric entangled state $|s\rangle$.

Fig. 6 (a) shows how the concurrence follows almost the same oscillations as the population of the symmetric entangled state, $\rho_{ss} =\frac{1}{\sqrt{2}}(|2\rangle + |3\rangle)$, for the same values as in Fig. 5, in particular,  $J=4\times10^9$ s$^{-1}$ and  $\Omega=4\times 10^7$ s$^{-1}$.  The frequency of the oscillations are close to the value of $\Omega$, being independent of the $J$ value provided $J$ be greater enough than $\Omega$ as is the case. The effect of $\gamma$ is manifest in the decay and in the anomalous behaviour in the minima of the concurrence, which can be appreciated in Fig. 6 (a).
In the ideal case of no decay,  the concurrence oscillates  exactly in the same manner as $\rho_{ss}$, the maxima of the concurrence corresponding to maximum population of the entangled state $|s\rangle$.

\begin{figure}
\centering
 \includegraphics[width=8cm]{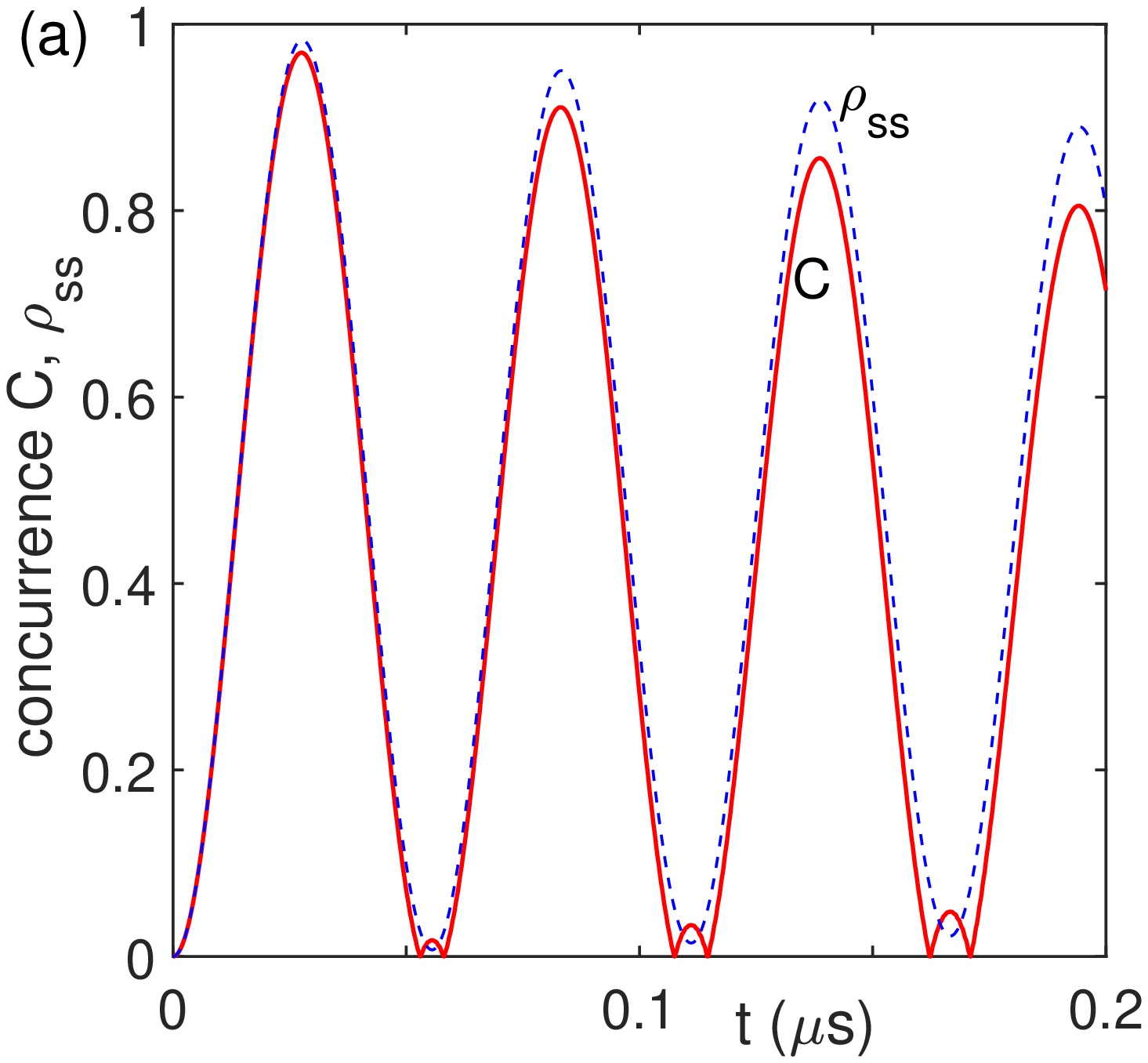}
 \includegraphics[width=8cm]{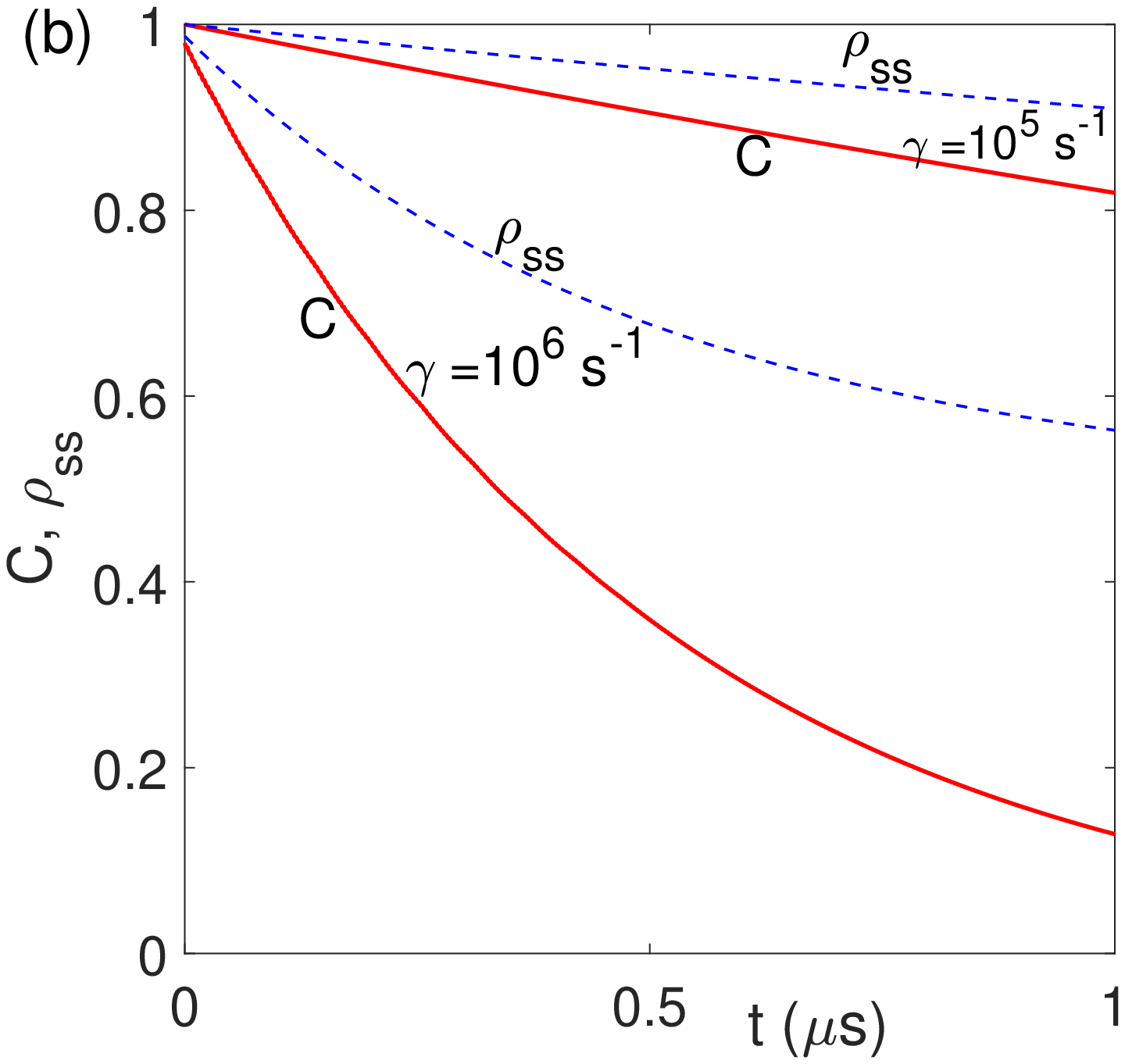}
 \caption{\label{Figure6}
(a) Temporal evolution of concurrence $C$ (red solid line) and  population $\rho_{ss}$ (blue dashed line)  for the initial state $|e_1 e_2\rangle$, $\Delta_l =J$, $\omega_0=1.5\times 10^{11}$ rad s$^{-1}$, $J=4\times 10^{9}$ s$^{-1}$, $\Omega=4\times 10^7$ s$^{-1}$ and  $\gamma=10^{6}$ s$^{-1}$.
  (b) Temporal evolution of concurrence $C$ (red solid lines) and population $\rho_{ss}$ (blue dashed lines)
  once the field is suppressed, for the same values as in (a), but $\gamma=10^{6}$ s$^{-1}$ in the two lower curves and $\gamma=10^{5}$ s$^{-1}$ in the two upper ones.}
  \end{figure}

This case offers the possibility to maintain the symmetric entangled state $|s\rangle$ for a time only limited by decoherence. It suffices to suppress the radiation field just when the population of this state  reaches a maximum, the concurrence being very close to unity. In the case shown in  Fig. 6 the first maximum occurs after $0.028$  $\mu$s from the initial time, but if  $\Omega$ decreases, the oscillations are slower, hence the first maximum of concurrence (and of $\rho_{ss}$) occurs latter.
Since the state $|s\rangle$ is eigenstate of the collective system and spontaneous emission is negligible, that entangled state can be kept without oscillations for long time if  decoherence  from environment can be minimized in high degree as it seems possible with the new technologies. In Fig. 6 (b), temporal evolution of the concurrence and population $\rho_{ss}$, once the field is suppressed,  are shown for $\gamma = 10^6$ s$^{-1}$ (the lower curves) and for $\gamma = 10^5$ s$^{-1}$ (the two upper curves with slower decay). In both cases, the temporal evolution of the coherence $Re \rho_{23}$ is found to be the same as that of $\rho_{ss}$.

The same result is obtained if the initial state is $|g_1g_2\rangle$ and the detuning $\Delta_l =-J$ since the state $|s\rangle$ is reached in analogous way as before.

C) If the initial state is $|e_1e_2\rangle$ and $\Delta_l=-J$, the resonance is established between the upper level and the entangled antisymmetric state $|a\rangle$, but it is a forbidden transition  and only a negligible  concurrence is obtained.

\section{Conclusions}

The entanglement between two non-planar and light identical molecules that present inversion doubling
of frequency $\omega_0$ in the microwave or far infrared range,
has been analyzed as a two two-level system with dipole-dipole interaction much lower than $\hbar \omega_0$.
The novelty of the work lies in the simplicity and peculiarities of this type of molecular system in which two near levels can be connected by allowed electric dipole transition with considerable electric dipole moment and negligible spontaneous emission because of the so small frequency $\omega_0$. The data used are those of the $NH_3$ molecule but other molecules  could  present the same advantageous features.  A coherence decay by pure dephasing induced by the environment has been assumed to be around $10^6$ s$^{-1}$.
If the  initial state is not eigenstate of the collective system, it evolves generating entanglement which oscillates in time with a frequency determined by the value of the dipole-dipole interaction $J$ (in the nanosecond range in our example)  and then with negligible coherence decay in this temporal range. Of particular interest are the generated entangled states $\frac{1}{\sqrt{2}}(|e_1g_2\rangle  \pm i |g_1e_2\rangle)$, whose populations oscillate between them with negligible decay, thus harmonically with frequency $J$. It would allow to preserve one of these entangled states  by quantum Zeno effect by frequent observations of such state  at time intervals lower that $J^{-1}$ s.

If the initial state is the collective lower eigenstate $|g_1g_2\rangle$ or upper eigenstate $|e_1e_2\rangle$, a driving coherent radiation field  of frequency  $\omega_0$ resonant with the inversion doubling and Rabi frequency $\Omega < J$,  generates oscillating entanglement by coherent superposition of these lower and upper states. This oscillation is slower than the Rabi oscillations and becomes  slower as the dipole-dipole interaction J grows. In our example, the oscillation period is
close to the $\mu$s range, hence, the coherence decay considered becomes manifest. With appropriate detuning of the field,
the symmetric entangled state  $|s\rangle = \frac{1}{\sqrt{2}}(|e_1g_2\rangle + |g_1e_2\rangle)$  is populated, its population oscillating with frequency close to the Rabi frequency $\Omega$ and independent on the $J$ value.
This entangled state, eigenstate of the collective system and with negligible spontaneous emission, could be preserved by  suppressing the field at appropriate time, its coherence being only limited by environmental decoherence processes which could be minimized according to technical  advances.
In order to consider less restrictive conditions, the effects of dissipative environments in this type of system will be explored in a future work.

Finally, concerning possible experiments, at least the symmetric top  $NH_3$ molecule can be easily prepared in particular states, as was done for the first maser operating just in the inversion doublet transition here considered. Such states correspond to the case of entanglement in presence of  an external driving field (Sec. 4).
To generate that entanglement, an oriented arrangement of molecules at appropriate distances is necessary, which can be achieved in particular optical lattices or arrays as proposed by several authors, e.g. \cite{Charron07,MOL3,MOL44,MOL5,MOL6}. Also, a low enough temperature is necessary, and although ultracold (less than 1 mK) polyatomic molecules are still rather difficult to prepare,
it is feasible in the current status to obtain cold samples (above mK), leading to entanglement coherence times of the order here shown. Then, it seems, in principle, that the mentioned entanglement in presence of an external driving field could be achievable shortly.

\section*{Conflicts of interest}

There are no conflicts to declare.

\section*{Acknowledgments}

We are very grateful to Miguel A. Porras and Fernando Carreno for technical help.
We acknowledge support from Project
FIS2017-87360-P, and I. G. acknowledges also support from Project FIS2016-76110-P, both projects from the Spanish Ministerio de Econom\'{\i}a y Competitividad (MINECO).


\begin{thebibliography}{99}

\bibitem{Mermin} N. D. Mermin, {\it Phys. Today}, {\bf 38} (4), 38 (1985).

\bibitem{DiVincenzo} D.  DiVincenzo,
{\it Fortschr. Phys.}, {\bf 48}, 771 (2000).

\bibitem{REV1} J. Singh and M. Singh.
Evolution in Quantum Computing.
 International Conference System Modeling and Advancement in Research Trends (SMART), 267 (2016).

\bibitem{BOOK1} E. Grumbling and M. Horowitz (Eds.). Quantum Computing: Progress and Prospects. National Academies of Sciences, Engineering, and Medicine. The National Academies Press. Washington, DC 2018. doi.org/10.17226/25196.

\bibitem{BOOK2} L. Jaeger. The Second Quantum Revolution: From Entanglement to Quantum Computing and Other Super-Technologies. Springer Nature Switzerland A G, 2018.


\bibitem{Bennett2} C. H. Bennett, G. Brassard, C. Crepeau, R. Jozsa, A. Peres and
W. K. Wootters, {\it Phys. Rev. Lett.}, {\bf  70}, 1895 (1993).

\bibitem{Gisin} N. Gisin, G. Ribordy, W. Tittel and H. Zbinden, {\it Rev. Mod. Phys.}, {\bf 74}, 145 (2002).

\bibitem{INF1} M. M. Wilde. Quantum Information Theory. Cambridge University Press, second edition 2017.


\bibitem{Horodecki} R. Horodecki, P. Horodecki, M. Horodecki and K. Horodecki,
{\it Rev. Mod. Phys.}, {\bf 81}, 865 (2009).

\bibitem{Ficek04} R. Tanas and Z. Ficek,
{\it J. Opt. B: Quantum  Semiclass. Opt.}, {\bf 6}, S90 (2004).

\bibitem{Ficek06} Z. Ficek and R. Tanas,
{\it Phys. Rev. A}, {\bf 74}, 024304 (2006).

\bibitem{Ficek10} Z. Ficek,
{\it Front. Phys. China}, {\bf 5} (1), 26 (2010).

\bibitem{Patrick10} S. R. J. Patrick, Fu-li Li, Meng Wang and Yang Yang,
{\it Journal of Modern Optics}, {\bf 57}, 4, 295 (2010).

\bibitem{Julsgaard12} B. Julsgaard and K. Molmer,
{\it Phys. Rev. A}, {\bf 85}, 032327 (2012).

\bibitem{Bashkirov14} E. K. Bashkirov and D. V. Litvinova,
Proc. SPIE 9448, Saratov Fall Meeting 2014: Optical Technologies in Biophysics and Medicine XVI; Laser Physics and Photonics XVI; and Computational Biophysics, 944827 (19 March 2015); doi: 10.1117/12.2179881.

\bibitem{Zhao17} Zhao Jin, Shi-Lei Su, Ai-Dong Zhu et al.,
{\it Scientific Reports}, {\bf 7} (1), 17648 (2017).


\bibitem{Charron07} E. Charron, P. Milman, A. Keller  and O. Atabek,
{\it Phys. Rev. A}, {\bf 75}, 033414 (2007).

\bibitem{MOL1} Q. Wei, S. Kais, B. Friedrich and D. Herschbach,
  {\it J. Chem. Phys.}, {\bf 135}, 154102 (2011).

\bibitem{MOL2} S. Kais (Ed.). Quantum Information and Computation for Chemistry. Adv. Chem. Phys. Vol 154, Wiley and Sons, Inc. 2014.

\bibitem{MOL3} M. Karra, K. Sharma, B. Friedrich, S. Kais, and D. Herschbach,
{\it J. Chem. Phys.} {\bf 144}, 094301 (2016).


\bibitem{MOL4} Z.-Y. Zhang and J-M. Liu,
Scientific Reports {\bf 7,} 17822 (2017).


\bibitem{Kotochigova06} S. Kotochigova1 and E. Tiesinga,
{\it Phys. Rev. A}, {\bf 73}, 041405(R) (2006).

\bibitem{MOL44} Q. Wei, S. Kais, and Y. P. Chen
{\it J. Chem. Phys.} {\bf 132}, 121104 (2010).

\bibitem{MOL5} Q. Wei, S. Kais, B. Friedrich and D. Herschbach,
{\it J. Chem. Phys.}, {\bf 134}, 124107 (2011).

\bibitem{MOL6} Q. Wei,Y. Cao, S. Kais, B. Friedrich and
    D.Herschbach,
{\it ChemPhysChem}, {\bf 17}, 1 (2016).


\bibitem{Grishanin05} B. Grishanin, H. Takahashi, Yu. Vladimirova, D. Zhdanov and V. N. Zadkov,
{\it Laser Physics}, {\bf 15}, 9,  1247 (2005).

\bibitem{Babilotte16} P. Babilotte, K. Hamraoui, F. Billard et al.,
{\it Phys. Rev. A}, {\bf 94}, 043403 (2016).

\bibitem{Carr09} L. D. Carr, D. DeMille, R. V. Krems and J. Ye,
{\it New J. Phys.}, {\bf 11}, 55049 (2009).

\bibitem{Jin12} D. S. Jin and J. Ye,
{\it Chem. Rev.}, {\bf 112}, 4801 (2012).

\bibitem{Park17} J. W. Park, Z. Z. Yan, H. Loh, S. A. Will and M. W. Zwierlein,
{\it Science}, {\bf 357}, Issue 6349, 372 (2017).

\bibitem{COL1} S. Truppe, H. J. Williams, M. Hambach, L. Caldwell, N. J. Fitch, E. A. Hinds, B. E. Sauer and M. R. Tarbutt,
{\it Nature Physics}, {\bf 13}, 1173 (2017).


\bibitem{Bader14} K. Bader, D. Dengler, S. Lenz Room  et al.,
{\it Nature Communications}, {\bf 5}, Article number: 5304 (2014).

\bibitem{Ferrando-Soria16} J. Ferrando-Soria, S. A. Magee, A. Chiesa et al.,
{\it Chem.}, {\bf 1}, 727 (2016).


\bibitem{Zureck} W. Zureck,
{\it Phys. Today}, {\bf 44}, 10, 36 (1991).

\bibitem{Isabel} I. Gonzalo in:
M. Ferrero and A. van der Merwe (eds.), New Deveiopments on Fundamental Problems in Quantum Physics. pp 59-162. KluwerAcademic Publishers 1997.

\bibitem{Jonalassinio} G. Jona-Lasinio, C. Presilla and C. Toninelli,
{\it Phys. Rev. Lett.}, {\bf 88}, 123001, (2002).

\bibitem{IsaBar2011} I. Gonzalo and P. Bargueno,
{\it Phys. Chem. Chem. Phys.}, {\bf 13}, 17130 (2011).

\bibitem{Coles}  Patrick J. Coles, Vlad Gheorghiu and Robert B. Griffiths,
{\it Phys. Rev.  A}, {\bf 86}, 042111 (2012).

\bibitem{Quack}  M. Quack, {\it Annu. Rev. Phys. Chem.}, {\bf 59}, 741 (2008).

\bibitem{Townes} C. H. Townes and A. L. Schawlow. Microwave Spectroscopy. McGrawHill, New York, 1955.

\bibitem{Zeno} B. Misra and E. C. G. Sudarshan, {\it J. Math. Phys. (N.Y.)}, {\bf 18}, 756 (1977).

\bibitem{Zeno2}P. Facchi and S. Pascazio,
{\it J. Phys. A: Math. Theor.}, {\bf 41}, 493001 (2008).

\bibitem{Zeno3} P. Facchi, S. Tasaki, S. Pascazio, H. Nakazato, A. Tokuse and D. A. Lidar,
{\it Phys. Rev. A}, {\bf 71}, 022302 (2005).

\bibitem{Liao10} Jie-Qiao Liao, Jin-Feng Huang, Le-Man Kuang and C. P. Sun,
{\it Phys. Rev. A}, {\bf 82}, 052109 (2010).

\bibitem{Wootters} W. K. Wootters, {\it Phys. Rev. Lett.}, {\bf 80}, 2245 (1998).
S. Hill  and W. K. Wootters, {\it Phys. Rev. Lett.}, {\bf 78}, 5022 (1997).

\bibitem{Feynmann} R. P. Feynman, R. B. Leighton and M. Sands. The Feynmann Lectures on Physics, Vol. 3, Chap. 9, Addison-Wesley, 1964.

\end{thebibliography}
\end{document}